\documentclass[12pt]{article}
\usepackage{graphicx}
\usepackage{amsmath}
\usepackage{siunitx}
\usepackage[UKenglish]{babel}
\usepackage{lipsum}
\usepackage{booktabs}
\usepackage{float}
\usepackage{xspace}
\usepackage{url}
\newcommand\pubnumber{ATL-PHYS-PROC-2018-120}
\newcommand\pubdate{\today}
\newcommand\ttbar{\ensuremath{t\bar{t}}\xspace}
\newcommand*{\ttV}{\ensuremath{\ttbar V}\xspace}
\newcommand\GeV{\ensuremath{\text{Ge\kern -0.1em V}}\xspace}

\newcommand\BLFV{\ensuremath{{\cal  B}(t \to \ell \ell' q)}\xspace}
\newcommand\BLFVta{\ensuremath{{\cal  B}(t \to e \mu q)}\xspace}

\newcommand\limexp{\ensuremath{1.36}\xspace}
\newcommand\limobs{\ensuremath{1.86}\xspace}
\newcommand\limexpnotau{\ensuremath{4.8}\xspace}
\newcommand\limobsnotau{\ensuremath{6.6}\xspace}
\newcommand*{\LFV}{cLFV\xspace}

\def\institute{Physikalisches Institut\\
Universit\"at Bonn}
\def\support{\footnote{Supported by European Research Council grant
ERC-CoG-617185.}}

\makeatletter
\def\blfootnote{\xdef\@thefnmark{}\@footnotetext}
\makeatother

\def\copyright{\blfootnote{Copyright 2018 CERN for the benefit of the ATLAS Collaboration. CC-BY-4.0 license.}}

\def\Title#1{\begin{center} {\Large #1 } \end{center}}
\def\Author#1{\begin{center}{ \sc #1} \end{center}}
\def\Address#1{\begin{center}{ \it #1} \end{center}}

\newcommand\pubblock{\rightline{\begin{tabular}{l} \pubnumber\\
         \pubdate  \end{tabular}}}
\newenvironment{Abstract}{\begin{quotation}  }{\end{quotation}}
\newenvironment{Presented}{\begin{quotation} \begin{center} 
             PRESENTED AT\end{center}\bigskip 
      \begin{center}\begin{large}}{\end{large}\end{center} \end{quotation}}

\begin{document}
\begin{titlepage}
\pubblock

\vfill
\Title{Search for charged lepton-flavour violation in top-quark decays at the LHC with the ATLAS detector}
\vfill
\Author{Carlo A. Gottardo\support, on behalf of the ATLAS Collaboration\copyright}
\Address{\institute}
\vfill
\begin{Abstract}
A direct search for charged lepton-flavour violation in top-quark decays is
presented. The data analysed correspond to $79.8\ \text{fb}^{-1}$ of
proton--proton collisions at a centre-of-mass energy of $\sqrt{s}=13\ 
\text{TeV}$ recorded by the ATLAS experiment at the LHC. The process studied is the
production of
top-quark pairs, where one top quark decays into a pair of opposite-sign
different-flavour charged leptons and an up-type quark, while the other decays
semileptonically according to the Standard Model. 
The signature of the signal is thus characterised by the presence of three
charged leptons, a light jet and a $b$-jet. A multivariate discriminant is
deployed and its distribution used as input to extract the signal strength.
In the absence of a signal, an upper limit on the
branching ratio of ${\cal  B}(t \to \ell \ell' q) < 1.86 \times 10^{-5}$ 
is set at the 95\% confidence level.
\end{Abstract}
\vfill
\begin{Presented}
$11^\mathrm{th}$ International Workshop on Top Quark Physics\\
Bad Neuenahr, Germany,  September 16--21, 2018
\end{Presented}
\vfill
\end{titlepage}
\def\thefootnote{\fnsymbol{footnote}}
\setcounter{footnote}{0}

\section{Introduction}
The observation of charged lepton-flavour violation (cLFV) would provide a
strong evidence for New Physics, beyond the simple inclusion of right-handed
neutrinos in the Standard Model (SM)~\cite{Calibbi:2017uvl}. 
Several extensions to the SM entail cLFV~\cite{Calibbi:2017uvl},
but a model-independent approach, based on effective field theory (EFT), is
adopted here. The set of operators reported in
Equation~\ref{eq:operators}, describing two-lepton, two-quarks contact
interactions, induces cLFV in processes involving the top quark.
The notation $e_{i} = \{e, \mu, \tau\}$ and
$u_{q} = \{u, c\}$ is used, and $H, H^\prime, \bar{H} = L, R$ 
indicate the chirality of the projectors $P$, with $H \neq \bar{H}$.
\begin{equation}
\label{eq:operators}
\begin{aligned}
& \mathcal{O}^{AV}_{HH^\prime} = (\bar{e}_{i}\gamma^{\alpha}P_{H}e_{j})(\bar{u}_{q}\gamma_{\alpha}P_{H^\prime}t), \\
& \mathcal{O}^{S+P}_{H} = (\bar{e}_{i}P_{H}e_{j})(\bar{u}_{q}P_{H}t), \\
& \mathcal{O}^{S-P}_{H} = (\bar{e}_{i}P_{H}e_{j})(\bar{u}_{q}P_{\bar{H}}t),\\
& \mathcal{O}^{LQ}_{H} = (\bar{u}_{q}P_{H}e_{j})(\bar{e}_{i}P_{H}t).
\end{aligned}
\end{equation}
The couplings of the operators are rather unconstrained, allowing for a
branching fraction of the decay $t \to e \mu q$, with $q=u,c$, of order $10^
{-3}$, well within the sensitivity of the Large Hadron
Collider (LHC) and not previously probed directly~\cite{Davidson:2015zza}.
Taking advantage of the copious production of top-quark pairs at the LHC, the
present analysis investigates the decay  $t (\bar{t}) \to \ell^{\pm} \ell^{'
\mp} q (\ell^{\pm} \ell^{' \mp} \bar{q})$ with $\ell = 
\{e,\mu, \tau \}$ and $q = \{ u,c\}$, in \ttbar events, where the other
top quark of the pair decays
semileptonically, according to the SM.
Signal events are generated at leading order precision in QCD with  
\textsc{MadGraph5}\_aMC@NLO interfaced with \textsc{Pythia}. The response of the
detector is simulated with \textsc{GEANT4}.
The search is performed using the $pp$ collision data collected by ATLAS~
\cite{atlas} in 2015, 2016 and 2017, corresponding to an integrated luminosity
of \SI{79.8}{\text{fb}^{-1}}.
More details about the analysis can be found in Reference~\cite{ATLAS:2018avw}.

\section{Analysis strategy}
Only events with three reconstructed charged light leptons (electrons or muons,
called leptons in the following) and at least two jets are selected.
Events having two opposite-sign same-flavour leptons whose invariant mass
is within $81.2$ and $101.2$ \GeV, or less then $15$ \GeV are vetoed.
At most one $b$-tagged jet is allowed. The signal region is defined by
requiring the presence of at least one electron and at least one muon. 
The signal kinematic is reconstructed, and a multivariate discriminant, namely a
boosted decision tree (BDT), is trained in the signal region in order to
separate the signal from all backgrounds. Thirteen input
variables are selected and provided to the BDT, which is trained on simulated
events.
Several sources of background populate the signal region: \ttbar, $Z+
\text{jets}$, diboson production, $t\bar{t}Z$, $t\bar{t}W$, $t\bar{t}H$,
associated single-top production as $tZ$, $tWZ$ and $tH$, and other minor
processes. 
Despite the fact that three isolated leptons are required, \ttbar and $Z+
\text{jets}$ constitute the largest fraction of background (60\% in the signal
region), due to the occurrence of additional non-prompt leptons
originating from secondary processes, such as the decay of a $b$- or $c$-hadrons,
photon conversion or object mis-reconstruction. This so-called non-prompt lepton
background is related to the details in the particles' interactions with the
detector material, therefore it has been modelled with a data-driven approach.
The technique used provides, through data events reweighting, both normalisation
and shape predictions for the non-prompt lepton background. The total expected
background is obtained by combining the data-driven non-prompt lepton background
prediction with simulated events containing three prompt leptons in the region's
acceptance.

Systematic uncertainties, including both normalisation and shape components,
are derived and assigned to the data-driven non-prompt lepton background.
Modelling and instrumental uncertainties are assigned to the simulated event
samples. The most impacting uncertainties are those associated to the non-prompt
lepton background, which correspond to a 10\% variation in the
total background yield and to a 9\% variation in the observed limit.

\section{Results}
\begin{figure}[H]
\begin{center}
\includegraphics[width=3.0in]{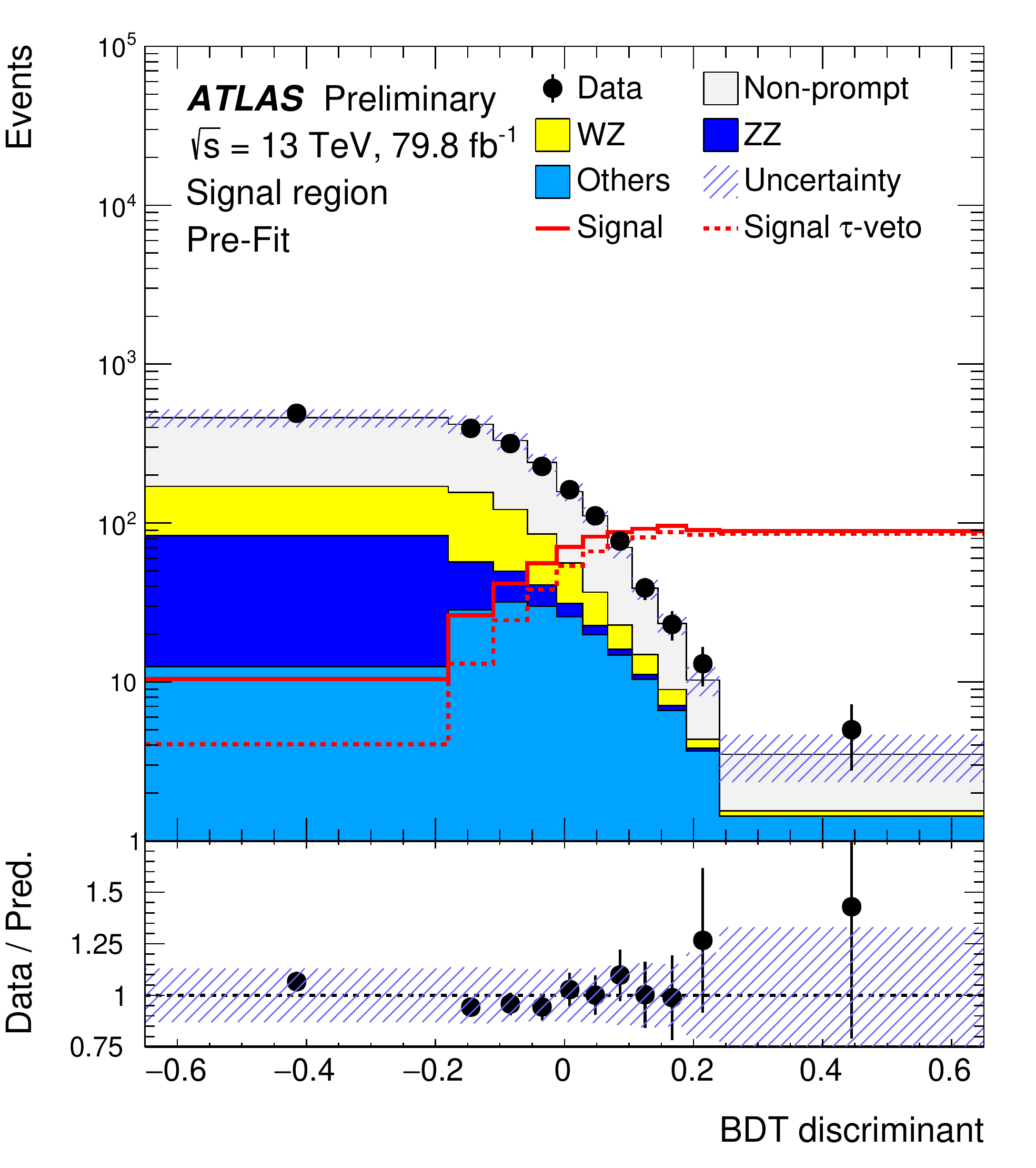}
\caption{BDT discriminant distribution, with the signal including and excluding
$\tau$ leptons (Signal $\tau$-veto) in the cLFV vertex overlaid. The signals are
normalised to the branching fraction ${\cal B}(t \to \ell^\pm \ell^
{\prime\mp}q)=3\times10^{-4}$ and ${\cal B}(t \to e \mu q)=1\times10^{-4}$,
respectively. All sources of systematic uncertainty are included.
Data (black points) are compared to the sum of backgrounds in the upper panel, 
while the ratio is shown in the lower panel~\cite{ATLAS:2018avw}.}
\label{fig:pre}
\end{center}
\end{figure}

The expected and observed event yields in the signal region are summarized in
Table~\ref{tab:prepost}.
The full BDT distribution, presented in Figure~\ref{fig:pre}, is used as
input for a binned profile-likelihood fit, meant to test for the presence of
signal events.
The data are found to be compatible with the absence of the signal and an upper
limit is set at the \SI{95}{\percent} confidence level, using the $CL_s$ method.
The observed and expected limits on the cLFV decay branching ratio are:
\[
    \BLFV < \limexp^{+0.61}_{-0.37} \times 10^{-5} \quad (\text{expected}),
\]
\[
    \BLFV < \limobs \times 10^{-5} \quad (\text{observed}).
\]
The upper limit is recomputed removing all generated signal events
where a $\tau$ lepton is present in the \LFV decay vertex, resulting in:
\[
    \BLFVta < \limexpnotau^{+2.1}_{-1.4} \times 10^{-6} \quad (\text{no $\tau$
    in \LFV
    vertex, expected}),
\]
\[
    \BLFVta < \limobsnotau \times 10^{-6} \quad (\text{no $\tau$ in \LFV vertex,
    observed}).
\]
\begin{table}[H]
\centering
\caption{Background and data events expected in the signal region~\cite{ATLAS:2018avw}.}
\medskip 
\begin{tabular}{c*{6}S}
\toprule
{Non-prompt} & $WZ$ & $ZZ$ & \multicolumn{1}{c}{\ttV} & {Other} & {Expected} & 
\multicolumn{1}{c}{Data} \\
{leptons}    &      &      &      & {prompt SM} & {events} & \\
\midrule
1190(180) & 350(140) & 140(52) & 108(10) & 76(10) & 1860(230) & \multicolumn{1}{c}{1857} \\
\bottomrule
\end{tabular}
\label{tab:prepost}
\end{table}

\begin{thebibliography}{99}

%%
%%  bibliographic items can be constructed using the LaTeX format in SPIRES:
%%    see    http://www.slac.stanford.edu/spires/hep/latex.html
%%  SPIRES will also supply the CITATION line information; please include it.
%%
%\cite{Aad:2008zzm}
%\cite{ATLAS:2018avw}

\bibitem{Calibbi:2017uvl} 
  L.~Calibbi and G.~Signorelli,
  %``Charged Lepton Flavour Violation: An Experimental and Theoretical Introduction,''
  Riv.\ Nuovo Cim.\  {\bf 41}, 1 (2018)
  [arXiv:1709.00294].
%%CITATION = PWASA,13,1564;%%

%\cite{Davidson:2015zza}
\bibitem{Davidson:2015zza}
  S.~Davidson, M.~L.~Mangano, S.~Perries and V.~Sordini,
  %``Lepton Flavour Violating top decays at the LHC,''
  Eur.\ Phys.\ J.\ C {\bf 75} (2015)  450
  [arXiv:1507.07163].
  %doi:10.1140/epjc/s10052-015-3649-5
  %%CITATION = doi:10.1140/epjc/s10052-015-3649-5;%%  

\bibitem{atlas}
  ATLAS Collaboration,
  %``The ATLAS Experiment at the CERN Large Hadron Collider,''
  JINST {\bf 3} (2008) S08003.\\
  %doi:10.1088/1748-0221/3/08/S08003
  %%CITATION = doi:10.1088/1748-0221/3/08/S08003;%%
  %6764 citations counted in INSPIRE as of 20 Sep 2018

\bibitem{ATLAS:2018avw}
  ATLAS Collaboration,
  %``Search for charged lepton-flavour violation in top-quark decays at the LHC with the ATLAS detector,''
  ATLAS-CONF-2018-044.\\
  \url{https://cds.cern.ch/record/2638305}
  %%CITATION = ATLAS-CONF-2018-044;%%

\end{thebibliography}
\end{document}